\documentclass[manuscript]{acmart}

\usepackage{url}
\renewcommand\footnotetextcopyrightpermission[1]{}

\AtBeginDocument{%
  \providecommand\BibTeX{{%
    \normalfont B\kern-0.5em{\scshape i\kern-0.25em b}\kern-0.8em\TeX}}}

\setcopyright{rightsretained}

\begin{document}

\title{Ten Challenges in Industrial Recommender Systems}

\author{Zhenhua Dong}
\email{dongzhenhua@huawei.com}
\author{Jieming Zhu}
\email{jamie.zhu@huawei.com}
\author{Weiwen Liu}
\email{liuweiwen8@huawei.com}
\author{Ruiming Tang}
\email{tangruiming@huawei.com}
\affiliation{%
  \institution{Huawei Noah’s Ark Lab}
  \city{Shenzhen}
  \country{China}
}

\renewcommand{\shortauthors}{Dong, et al.}



\maketitle
Huawei’s vision and mission is to build a fully connected intelligent world. Since 2013, Huawei Noah’s Ark Lab has helped many products to build recommender systems and search engines for getting the right information to the right users. Every day, our recommender systems serve hundreds of millions of mobile phone users and recommend different kinds of content and services such as apps, news feeds, songs, videos, books, themes, and instant services. The big data and various scenarios provide us with great opportunities to develop advanced recommendation technologies. Furthermore, we have witnessed the technical trend of recommendation models in the past ten years, from the shallow and simple models like collaborative filtering, linear model, low rank models to the deep and complex models like neural network, pre-trained language models. Based on the mission, opportunities and technological trends, we have also met several hard problems in our recommender systems. In this talk, we will share ten important and interesting challenges and hope that the RecSys community can get inspired and create better recommender systems.

Challenge 1 Missing information. The performance of recommendation models largely depend on the completeness and accuracy of training data. Unfortunately, the collected data from real industrial recommender systems is usually incomplete. We may miss both features (column data) and samples (raw data). For missing features, 
a user would like to watch a movie because of her friend’s suggestion, while the system probably does not collect that causal feature. For missing samples, recommender systems can only expose a few items to a user at one time, so we have no chance to know the user’s preferences on other items. It is important and challenging to mitigate the negative effect of missing data on the model.

Challenge 2 Individual treatment effect. The same feature always has a different causal effect (or influence) on different users’ preferences and decisions (e.g., click). For example, both user A and user B rate one movie 5 stars, while A likes the movie for the actors and B likes it for the director. The feature effect on the preference is different. How can we compute the individual treatment effect in recommender systems? The answer to this research question can help us to find the user’s causal features to an item and build the user interest preferences accurately. In \cite{pmlr-v216-zhao23a}, we proposed a conditional counterfactual causal effect (CCCE) method for individual treatment effect computation. CCCE can make use of the individual observation as evidence and find the causal attribution under certain proper assumptions. However, it is still a primary work, and we need to find more powerful methods to compute individual treatment effects under more generalized assumptions.

Challenge 3 Biases. There are many biases \cite{wu2022opportunity} in recommender systems due to missing data, confounders, and close feedback loops. Both academia and industry have done a lot of work about analyzing these biases and creating many debiasing methods \cite{dong2020counterfactual}, such as causal technologies and representation learning. However, there are still many hard problems that need to be handled. On the one hand, we find some new biases such as trust bias in ranking \cite{Zhao2023TrustBias}, duration bias in short video \cite{Zhao2023UncoveringUI}, and previous model bias; on the other hand, we need more powerful methods to solve the three important research questions such as how to collect unbiased data? How to train an unbiased model? How to evaluate the biases?

Challenge 4 Models resusing. Most of the current industrial recommendation models need to be re-trained with recent data for better performance, while the historical models are always underutilized, so it wastes a lot of computing resources. It is important that how to efficiently reuse these historical recommendation models. In \cite{Wang2023DataFree}, we propose a model inversed data synthesis framework to recover fewer representative samples from the historical models. Then, we can combine the recovered samples and the recent samples to train more powerful CTR prediction models. However, this is still a primary work, and more novel methods are needed to reuse the historical recommendation models efficiently and effectively.

Challenge 5 Large language model enhanced recommendation. ChatGPT has demonstrated the great capabilities of large language models. It is natural to think about how to make use of the advantages of LLMs to enhance recommender systems. In the survey \cite{lin2023recommender}, we summarize the current methods from two perspectives: where and how. We also highlight key challenges such as training efficiency, inference latency, ID indexing and modeling, user behavior text modeling, and biases in LLMs. LLMs are coming, and our recommender system communities should embrace this novel technology. As a poem says:
It does not matter if you love it or not, it is standing right there with no emotion, not going to change.

Challenge 6 Multiple modalities. Current mainstream recommendation models are mostly built on user behavior features such as historically interaction items and categories. These features naturally capture the collaborative signals from user behavior data, which better memorizes and learns over the co-occurrence patterns of the behavior features. However, the learning of feature embeddings is well known to be prone to overfitting and cannot generalize well to long-tailed and cold-start items. On the other hand, with the prevalence of large pre-trained models in computer vision (CV) and natural language processing (NLP), it has become more and more popular to employ pre-trained models to extract multimodal content features for improving recommendation performance, which is helpful for long-tailed or cold-start items. However, most of the large pre-trained models are trained for CV and NLP tasks, which are strong for semantic understanding but weak in co-occurrence pattern reasoning. This gap between content semantic understanding and user behavior-based collaborative filtering leads to minor improvements or sub-optimal performance in recommendation. In a unified perspective, user behavior sequences can be viewed as another modality along with content modalities such as texts and images. It remains an open challenging problem to build a recommendation-focused multi-modal pre-trained model that can bridge user behavior modality and content modalities.

Challenge 7 Simulation. User modeling is a classical research topic, with the UMAP conference having a history of more than 30 years \cite{UMAPHistory}. Many great research questions have been well studied, but as we know, since people are complex and recommender systems often miss a lot of information, it is impossible to know everything about a user, making it quite difficult to accurately model their preferences. In many advertising scenarios, the CTR accuracy is less than 1\%. Faced with this challenge, researchers from Renmin University of China have proposed a novel paradigm called RecAgent \cite{wang2023recagent}, which is an LLM-based recommendation simulator. RecAgent can simulate user behaviors such as browsing, chatting, broadcasting, and consuming recommended items. It is a digital twin for recommender systems and has the potential to align simulated user behaviors with real human understanding. Although RecAgent is impressive, simulation for people and recommendation is still very challenging. Questions remain about how to simulate more people and user behaviors efficiently and accurately, and how to evaluate the simulation?

Challenge 8 Lifetime value (LTV) modeling. Most of current recommender systems only optimize the short-term objectives such as click, rating and dwell time, which can not align the goal of recommender systems in improving the long-term user experiences. The user's feedback about long-term satisfaction is usually sparse, volatile and delayed, especially for the deep conversion tasks, like lifetime value. In the tutorial \cite{LTVTutorial2023}, we introduce the definitions and scenarios of LTV, some typical LTV prediction technique and our practice in products. While there are still many hard problems about LTV modeling, such as delayed feedback, cold start, offline evaluation and multi-task optimizations.

Challenge 9 Trustworthy. Recommender systems should take responsibility for social goodness. We think that trustworthy recommender systems should consider eight perspectives, such as accountability, transparency, robustness, privacy, security, fairness, assisting or serving people, and long-term enhancement of the happiness of humans, society and the environment. Most current recommender systems focus on short-term optimization such as CTR and watch time in one interaction. However, it is not enough to be a trustworthy recommender system. We really hope that more scholars can help the industry to define and build trustworthy and socially good recommender systems from different perspectives such as society, economy, user-centric, ecological systems and natural environments.

Challenge 10 Win-win ecosystem.
 There are mainly four important stakeholders in an information system: users, content provider (CP), advertisers, and platforms. A dialog-based information system is good at providing direct answers to users, but this method may hurt the benefits of some stakeholders. Here are some key considerations for each stakeholder:
1.	For content providers: How to protect their intellectual property and benefits? The traditional list-based content exposing method can lead users to click on the landing page of the CP, where the CP can expose its advertisements.
2.	For advertisers: How can advertisements be appropriately exposed during the dialog process?
3.	For users: How can we ensure that the generated information is objective and accountable?
4.	For search and recommender systems: How to design win-win interactions and mechanisms for long-term benefits?

Besides the above ten challenges, there are still a lot of ones in recommender systems, we would like to share novel research questions and offer collaborative opportunities and resources with academia. More interesting and technical challenges from Huawei can be found in ChaSpark \cite{chaSpark2023}.
Let us think about the future of recommendation services, based on the emergence of more and more technological innovations, such as large language models, multi-modality, causality etc. We believe that there will be more and more intelligent user-centric services. In \cite{dong2022brief}, we discuss some potential services of the future recommender system, a British butler, who serve master well based on his understanding and long serving experiences; a knowledgeable teacher, who can educate us about new knowledge and useful information for our lifelong growth. Moreover, the future recommender systems have chance to be Samantha in “Her”, Baymax in “Big Hero 6”, JARVIS in “Iron Man”. The vision is great, while there is a huge gap between the vision and the reality, and the gap means great research opportunities.



\begin{acks}
Thanks to the colleagues from Huawei Noah’s Ark Lab and product team.
\end{acks}

\bibliographystyle{ACM-Reference-Format}
\bibliography{reference}

\end{document}